\documentclass[twocolumn,showpacs,preprintnumbers,amsmath,amssymb]{revtex4}

\usepackage{graphicx}% Include figure files
\usepackage{dcolumn}% Align table columns on decimal point
\usepackage{bm}% bold math
\usepackage{mathrsfs}
\usepackage{color}

\begin{document}

\title{Consistent description of $^{12}$C and $^{16}$O
using finite range three-body interaction
}% 

\author{N. Itagaki}

\affiliation{
Yukawa Institute for Theoretical Physics, Kyoto University,
Kitashirakawa Oiwake-Cho, Kyoto 606-8502, Japan
}

\date{\today}

\begin{abstract}
Consistent description of $^{12}$C and $^{16}$O has been
a long standing problem of microscopic $\alpha$ cluster models, where
the wave function is fully antisymmetrized and the effective interaction
is applied not between $\alpha$ clusters but between nucleons.
When the effective interaction is designed to reproduce the
binding energy of $^{16}$O (four $\alpha$), the binding energy of $^{12}$C 
(three $\alpha$) becomes underbound by about 10 MeV.
In the present study, by taking into account the coupling with the $jj$-coupling shell model
components and utilizing Tohsaki interaction, which is phenomenological but has
finite-range three-body interaction terms, we show that
consistent understanding of these nuclei can be achieved.
The original Tohsaki interaction gives small overbound of about 3 MeV for $^{16}$O,
and this is improved by slightly modifying three-body Majorana exchange 
parameter. Also, the coupling with the $jj$-coupling shell model
wave function strongly contributes to the increase of the binding energy of $^{12}$C.
So far the application of Tohsaki interaction has been limited
to $4N$ nuclei, and here, we add Bartlet and Heisenberg exchange terms
in the two-body interaction for the purpose of applying it to neutron-rich systems,
and it is applied to $^6$He.
\end{abstract}

\pacs{21.30.Fe, 21.60.Cs, 21.60.Gx, 27.20.+n}% PACS, the Physics and Astronomy
                             % Classification Scheme.
%\keywords{Suggested keywords}%Use showkeys class option if keyword
                              %display desired
\maketitle

\section{Introduction}

The nuclei $^{12}$C and $^{16}$O are typical light nuclei, and they 
have been extensively studied based on the cluster approaches~\cite{Fujiwara}. 
Since there is no bound nucleus with mass number 5 or 8,
formation of  $^{12}$C from three $^4$He nuclei ($\alpha$ clusters) is a key process
of the nucleosynthesis. Here, the second $0^+$ state
at $E_x = 7.6542$ MeV plays a crucial role, which is 
the second excited state of $^{12}$C and located
just above the threshold energy  to decay into three $^4$He nuclei~\cite{Hoyle}.
The existence of three $\alpha$ state just at this energy is
essential factor in the synthesis of various elements in starts.
For also $^{16}$O, cluster structure has been shown to be extremely important.
Although the ground state corresponds to the doubly closed $p$ shell of the shell model,
this configuration can be also interpreted from four $\alpha$ and $^{12}$C+$\alpha$ view points,
if we take certain limit for the inter cluster distances.
Also the first excited state of $^{16}$O at $E_x = 6.0494$ MeV, very close to the
threshold to decay into $^{12}$C and $^4$He,
can be interpreted as $^{12}$C+$^4$He cluster state~\cite{Horiuchi}, 
and low-lying cluster states just around the threshold are quite
important in the synthesis of $^{16}$O in starts~\cite{Descouvemont}.
Various cluster models have been proposed and successfully applied to these nuclei.

However consistent description of $^{12}$C and $^{16}$O has been
a long standing problem of microscopic $\alpha$ cluster models.
Here the definition of the microscopic model is that
the wave function is fully antisymmetrized and the effective interaction
is applied not between $\alpha$ clusters but between nucleons.
When the effective interaction is designed to reproduce the
binding energy of $^{16}$O (four $\alpha$), the binding energy of $^{12}$C (three $\alpha$)
becomes underbound by about 10 MeV, and on the contrary,
when the binding energy of $^{12}$C is reproduced,
$^{16}$O becomes overbound by about 20 MeV.
We have previously utilized Tohsaki interaction~\cite{Tohsaki}, which has
finite-range three-body terms, and the obtained result is better
than ones only within the two-body terms; however the problem
has not been fully solved~\cite{C-O}.

One of the clue to solve this problem is the inclusion of the spin-orbit interaction.
In most of the cluster models, $\alpha$ clusters are defined as 
simple $(0s)^4$ configuration at some point. 
These $\alpha$ clusters are spin singlet systems and the spin-orbit interaction
does not contribute inside $\alpha$ clusters and also between $\alpha$ clusters.
In $jj$-coupling shell model, the spin-orbit interaction is quite important and
this plays an essential role in explaining the observed magic numbers.
According to the $jj$-coupling shell model, $^{12}$C corresponds to the
subclosure configuration of spin-orbit attractive orbits ($p_{3/2}$) 
and the spin-orbit interaction works attractively, whereas $^{16}$O
corresponds to the closure of major shell ($p$ shell), where 
both spin-orbit attractive ($p_{3/2}$)  and repulsive ($p_{1/2}$) orbits
are filled and the contribution of the spin-orbit interaction cancels. 
Therefore, inclusion of the $\alpha$ breaking wave function and
taking into account the spin-orbit contribution are expected to
decrease the binding energy difference of $^{12}$C and $^{16}$O.
To describe the $jj$-coupling shell model states
with the spin-orbit contribution starting with the cluster model wave function,
we proposed the antisymmetrized quasi-cluster model 
(AQCM)~\cite{Simple,Masui,Yoshida2,Ne-Mg,Suhara,Suhara2015,Itagaki}.
In the AQCM, the transition from the cluster- to shell-model-structure 
can be described by only two parameters: $R$ representing the distance between $\alpha$ clusters
and $\Lambda$, which characterizes the transition of $\alpha$ cluster(s) to quasi-cluster(s) 
and quantifies the role of the spin-orbit 
interaction.

In nuclear structure calculations, it is quite well known that
the central part of the nucleon-nucleon interaction in the calculation  
should have proper density dependence in order to
to satisfy the saturation property of nuclear systems. 
If we just introduce simple
two-body interaction, for instance Volkov interaction~\cite{Volkov}, which has been
widely used in the cluster studies, we have to
properly choose Majorana exchange parameter for each nucleus, and consistent description of
two different nuclei with the same Hamiltonian becomes a tough work.
Thus it is rather difficult to 
reproduce the threshold energies to decay into different subsystems.

Concerning the density dependence of the interaction,
adding zero range three-body interaction term helps better agreements with
experiments as in modified Volkov (MV) interaction ~\cite{MV1}.
However, in this case the binding energies become quite sensitive to the 
choice of the size parameter of Gaussian-type single particle wave function.
Especially, the binding energy and radius of $^4$He cannot be reproduced consistently.
This situation is essentially common in the case of Gogny interaction widely used in
mean field studies~\cite{Gogny}. 
The nucleus $^4$He is a building block of $\alpha$ cluster states and it is desired that
the size and binding energy are reproduced. 
The Tohsaki interaction, which has finite range three-body terms, has much advantages
compared with the zero range three-body interactions. 
This interaction is a phenomenological one and 
designed to reproduce the $\alpha$-$\alpha$ scattering phase shift.
Also it gives reasonable size and binding energy of the $\alpha$ cluster,
which is rather difficult in the case of zero-range three-body interaction, and
the binding energy is less sensitive to the choice of size parameter of Gaussian-type
single particle wave function
because of the finite range effect of the three-body interaction.
Furthermore, the saturation properties of nuclear matter is reproduced rather satisfactory.

Of course,
introducing the term proportional to the fractional 
power of the density is 
another possibility to reproduce the saturation properties of nuclear systems
as in density functional theories (DFT),
instead of introducing three-body interaction terms.
However, here we perform parity and angular momentum projections, and we also
superpose many Slater determinants based on the generator coordinate method (GCM).
In this case, it is desired that the Hamiltonian is expressed in a operator form such
as three-body interaction, which enables us to calculate the transition matrix elements
between different Slater determinants.
From this view point, simplified version of finite range 
three-body interaction is proposed in Ref.~\cite{Enyo}.

The purpose of the present work is to combine the use of finite range three-body 
interaction for the interaction part
and AQCM for the wave function part
to establish consistent understanding of $^{12}$C and $^{16}$O. 
The original Tohsaki interaction gives small overbound for $^{16}$O (about 3 MeV)~\cite{C-O},
and here, we try to improve by slightly modifying three-body Majorana exchange 
parameter. For $^{12}$C, subclosure configuration of $jj$-coupling shell model,
where the spin-orbit interaction plays an important role, is coupled to
three $\alpha$ model spaced based on AQCM. For $^{16}$O, the closed shell 
configuration of the $p$ shell is the dominant configuration of the ground
state, and we apply four $\alpha$ model, which coveres the model space 
of closed $p$ shell.
Also, the application of Tohsaki interaction has been limited
to $4N$ nuclei, and we add Bartlet and Heisenberg exchange terms
in the two-body interaction for the purpose of applying it to neutron-rich systems.

% The paper is organized as follows. 
% The formulation is given in Sect.~\ref{model}. 
% In Sect.~\ref{results}, the AQCM results are shown. 
% Finally, in Sect.~\ref{summary} we summarize the results and give the main conclusions.

\section{The Model\label{model}}

\subsection{Hamiltonian}

The Hamiltonian ($\hat{H}$) consists of kinetic energy ($\hat{T}$) and 
potential energy ($\hat{V}$) terms,
\begin{equation}
\hat{H} = \hat{T} +\hat{V},
\end{equation}
and the kinetic energy term is described as one-body operator,
\begin{equation}
\hat{T} = \sum_i \hat{t_i} - T_{cm},
\end{equation}
and the center of mass kinetic energy ($T_{cm}$),
which is constant,
is subtracted.
The potential energy has
central ($\hat{V}_{central}$), spin-orbit ($\hat{V}_{spin-orbit}$), 
and the Coulomb parts.

\subsection{Tohsaki Interaction}
For the central part of the potential energy
($\hat{V}_{central}$), we adopt Tohsaki interaction.
The Tohsaki interaction consists of two-body ($V^{(2)}$)  and three-body ($V^{(3)}$) terms:

\begin{equation}
\hat{V}_{central} = {1 \over 2} \sum_{ij} V^{(2)}_{ij} 
+ {1 \over 6} \sum_{ijk}  V^{(3)}_{ijk},
\end{equation}
where $V^{(2)}_{ij}$ and $V^{(3)}_{ijk}$ consist of three terms,
\begin{equation}
V^{(2)}_{ij} =  
\sum_{\alpha=1}^3 V^{(2)}_\alpha \exp[- (\vec r_i - \vec r_j )^2 / \mu_\alpha^2]
 (W^{(2)}_\alpha + M^{(2)}_\alpha P^r)_{ij},
\label{2body}
\end{equation} 
\begin{eqnarray}
V^{(3)}_{ijk} = 
\sum_{\alpha=1}^3 && V^{(3)}_\alpha \exp[- (\vec r_i - \vec r_j )^2 / \mu_\alpha^2 -
                                                     (\vec r_i - \vec r_k)^2 / \mu_\alpha^2   ] \nonumber \\
\times &&  (W_\alpha^{(3)} + M_\alpha^{(3)} P^r)_{ij} (W_\alpha^{(3)} + M_\alpha^{(3)} P^r)_{ik}.
\end{eqnarray}
Here, $P^r$ represents the exchange of spatial part of the wave functions
of interacting two nucleons, and this is equal to $-P^\sigma P^\tau$ due to the Pauli principle
($P^rP^\sigma P^\tau = -1$),
where $P^\sigma$ and $P^\tau$ are spin and isospin exchange operators, respectively.
The range parameters $\{ \mu_\alpha \}$ are set to be common for the two-body and three-body parts,
and the values are listed in Table~\ref{Tohsaki-2} together with strengths of 
two-body interaction $\{ V^{(2)}_\alpha \}$, three-body interaction $\{ V^{(3)}_\alpha \}$, and
the Majorana exchange parameters of the two-body interaction.
The values of Winger parameters, $\{W^{(2)}_\alpha \}$, are given as
$W^{(2)}_\alpha = 1 - M^{(2)}_\alpha$.  We employ F1 parameter set in Ref.~\cite{Tohsaki}.

Until now Tohsaki interaction has been applied only to $4N$ nuclei,
and in this article we extend the application to neutron-rich nuclei.
The original Tohsaki interaction does have Wigner and Majorana exchange terms,
but spin and isospin exchange terms are missing. 
Because of this, the interaction gives a weak bound state for a two neutron system,
as in the case of Volkov interaction.
Therefore,
here we add Bartlet ($BP^\sigma$) and Heisenberg ($HP^\tau$) exchange terms in Eq.~\ref{2body} as
$(W^{(2)}_\alpha + BP^\sigma - HP^\tau + M^{(2)}_\alpha P^r)_{ij}$. 
The values of $B$ and $H$ are chosen to be 0.1. 
By adding these terms, the neutron-neutron interaction (or pron-proton interaction)
because weaker than the original interaction, while $\alpha$-$\alpha$ scattering phase shift
is not influenced by this modification.

\begin{table} 
 \caption{ 
Parameter set for the two-body part of the Tohsaki interaction
(F1 parameterization in Ref.~\cite{Tohsaki}) together with the strengths
of the three-body interaction.}
  \begin{tabular}{cccccc} \hline \hline
 $\alpha$  & $\mu_\alpha$ (fm)  & $V^{(2)}_\alpha$ (MeV) & $V^{(3)}_\alpha$ (MeV) & $M^{(2)}_\alpha$ & $W^{(2)}_\alpha$ \\ \hline
   1  &  2.5 & $-5.00$  & $-0.31$ & 0.75 & 0.25 \\  
   2  &  1.8 & $-43.51$&  7.73  & 0.462 & 0.538 \\  
   3  &  0.7 & $60.38$ &  219.0 & 0.522 & 0.478 \\ \hline 
  \end{tabular}   \\ 
\label{Tohsaki-2}
\end{table}

\begin{table} 
 \caption{ 
Majorana exchange parameters for the three-body interaction terms.
F1 stands for the original F1 set of Tohsaki interaction~\cite{Tohsaki},
and F1' is the modified versions introduced in the present article.
}
  \begin{tabular}{ccc} \hline \hline
 & F1 & F1'  \\ \hline
   $M^{(3)}_1$  &  0.0     & 0.0  \\ 
   $M^{(3)}_2$  &  0.0     & 0.0  \\ 
   $M^{(3)}_3$  &  1.909 & 1.5    \\ \hline 
  \end{tabular}   \\ 
\label{Tohsaki-3}
\end{table}

\subsection{Spin-orbit interaction}
For the spin-orbit part,
G3RS \cite{G3RS}, which is a realistic
interaction originally determined to reproduce the nucleon-nucleon scattering phase shift, 
is adopted;
\begin{equation}
\hat{V}_{spin-orbit}= {1 \over 2} \sum_{ij} V^{ls}_{ij}
\end{equation}
\begin{equation}
V^{ls}_{ij}= V_{ls}( e^{-d_{1} (\vec r_i - \vec r_j)^{2}}
                    -e^{-d_{2} (\vec r_i - \vec r_j)^{2}}) 
                     P(^{3}O){\vec{L}}\cdot{\vec{S}},
\label{Vls}
\end{equation}
 where $d_{1}= 5. 0$ fm$^{-2},\ d_{2}= 2. 778$ fm$^{-2}$,
and $P(^{3}O)$ is a projection operator onto a triplet odd state.
The operator $\vec{L}$ stands for the relative angular momentum
 and $\vec{S}$ is the total spin, ($\vec{S} = \vec{S_{1}}+\vec{S_{2}}$).
The strength,
$V_{ls}$, has been determined to reproduce the $^4$He+$n$
scattering phase shift~\cite{Okabe}, and
$V_{ls} = 1600-2000$ MeV has been suggested.

\subsection{Single particle wave function (Brink model)}

In conventional $\alpha$ cluster models, 
the single particle wave function has a Gaussian shape \cite{Brink};
\begin{equation}
	\phi_{i} = \left( \frac{2\nu}{\pi} \right)^{\frac{3}{4}}
	\exp \left[- \nu \left(\bm{r}_{i} - \bm{R}_i \right)^{2} \right] \eta_{i},
\label{Brink-wf}
\end{equation}
where $\eta_{i}$ represents the spin-isospin part of the wave function, 
and $\bm{R}_i$ is a real parameter representing the center of a Gaussian 
wave function for the $i$-th particle. 
In this Brink-Bloch wave function, four nucleons in one $\alpha$ cluster share the common $\bm{R}_i$ value. 
Hence, the contribution of the spin-orbit interaction vanishes. 

\subsection{Single particle wave function in the AQCM}

In the AQCM, $\alpha$ clusters are changed into quasi clusters. 
For nucleons in the quasi cluster, 
the single particle wave function is described by 
a Gaussian wave packet, and
the center of this packet $\bm{\zeta}_{i}$ is a complex parameter;
\begin{equation}
	\psi_{i} = \left( \frac{2\nu}{\pi} \right)^{\frac{3}{4}}
		\exp \left[- \nu \left(\bm{r}_{i} - \bm{\zeta}_{i} \right)^{2} \right] \chi_{i} \tau_{i}, 
\label{AQCM_sp} \\
\end{equation}
where
$\chi_{i}$ and $\tau_{i}$ in Eq.~\eqref{AQCM_sp} represent the spin and isospin parts of the $i$-th 
single particle wave function, respectively. 
The spin orientation is governed by the parameters 
$\xi_{i\uparrow}$ and $\xi_{i\downarrow}$,
which are in general complex, 
while the isospin part is fixed to 
be `up' (proton) or `down' (neutron),
\begin{equation}
	\chi_{i} = \xi_{i\uparrow} |\uparrow \ \rangle + \xi_{i\downarrow} |\downarrow \ \rangle,\\
	\tau_{i} = |p \rangle \ \text{or} \ |n \rangle.
\end{equation}
The center of Gaussian wave packet is give as
\begin{equation}
\bm{\zeta}_{i} = \bm{R}_i + i \Lambda \bm{e}^{\text{spin}}_{i} \times \bm{R}_i, 
\label{center}
\end{equation}
where $\bm{e}^{\text{spin}}_{i}$ is a unit vector for the intrinsic-spin orientation, 
and $\Lambda$ is a real control parameter describing the dissolution of the $\alpha$ cluster. 
As one can see immediately, the $\Lambda = 0$ AQCM wave function, which has no imaginary part, 
is the same as the conventional Brink-Bloch wave function.
The AQCM wave function corresponds to the $jj$-coupling shell model wave function,
such as subshell closure configuration,
when $\Lambda = 1$ and $\bm{R}_i \rightarrow 0$.
The mathematical explanation for this is summarized in Ref.~\cite{Suhara}.
For the width parameter, we use the value of $\nu = 0.23$ fm$^{-2}$.

\subsection{AQCM wave function of the total system}

The wave function of the total system $\Psi$ is antisymmetrized product of these
single particle wave functions;
\begin{equation}
\Psi = {\cal A} \{ \psi_1 \psi_2 \psi_3 \cdot \cdot \cdot \cdot \psi_A \}.
\label{total-wf}
\end{equation}  
The projections onto parity and angular momentum eigen states can be performed by 
introducing the projection operators $P^J_{MK}$ and $P^\pi$,
and these are performed
numerically in the actual calculation.

\subsection{Superposition of different configurations}

Based on GCM, the superposition of
different AQCM wave functions can be done,
\begin{equation}
\Phi = \sum_i c_i P^J_{MK} P^\pi \Psi_i.
\label{GCM}
\end{equation}
Here, $\{ \Psi_i\}$ is a set of AQCM wave functions with
different values of the $R$ and $\Lambda$ parameters,
and the coefficients for the linear combination, $\{ c_i \}$ are
obtained by solving the Hill-Wheeler equation~\cite{Brink}.
We obtain a set of coefficients for the linear combination $\{c_j\}$ for
each eigen value of $E$.

\section{Results}

\begin{figure}[t]
	\centering
	\includegraphics[width=6.0cm]{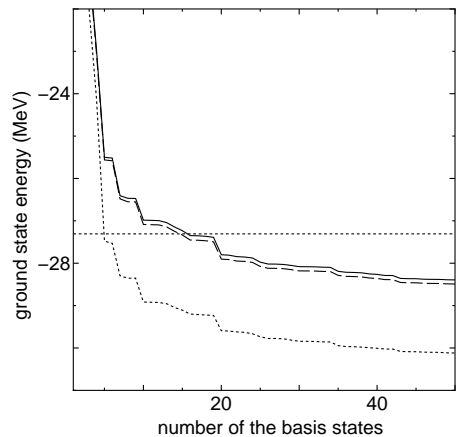} 
	\caption{
The energy convergence of 
the ground $0^+$ state of $^{6}$He with
an  $\alpha$+$n$+$n$ model.
We prepare different 
configurations for the two neutrons
outside $^4$He and superpose them. 
The horizontal line at $-27.31$ MeV
shows the threshold energy of $^4$He+$n$+$n$.
The dotted line is the result of original F1 parameter set, 
and the dashed line is the  result after adding $B=H=0.1$.
The results of F1' 
parameter sets is shown as the solid line.
     }
\label{he6conv}
\end{figure}

We start the application of Tohsaki interaction with $^6$He.
So far Tohsaki interaction has been applied to 4$N$ nuclei, and here
a neutron-rich system is examined.
Figure \ref{he6conv} shows the energy convergence of 
the ground $0^+$ state of $^{6}$He.
The model space is $\alpha$+$n$+$n$ and the positions of two
neutrons are randomly generated, and we superpose different 
configurations for the neutrons. The horizontal line at $-27.31$ MeV
shows the threshold energy of $^4$He+$n$+$n$
(the experimental value is $-28.29566$ MeV).
Here, the dotted line is the result of original F1 parameter set.
We add the Bartlet and Heisenberg terms with the parameters of 
$B = H = 0.1$, and the result is shown as the dashed line. 
Furthermore, we slightly modify the Majorana exchange parameter
of the three-body interaction term and the results of F1'
parameter set is shown as solid line
(this is to reproduce the binding energy of $^{16}$O,
which will be discussed shortly).
For the spin-orbit part,
the strength 
$V_{ls}$ has been determined to be $1600-2000$ MeV 
in the analysis of $^4$He+$n$
scattering phase shift~\cite{Okabe}, and here we adopted $V_{ls} = 1600$ MeV.
The results of dashed and solid lines are similar and
difficult to distinguish. Nevertheless, by adding Bartlet and Heisenberg
terms, we obtained the binding energy close to the experimental onei
(the experimental value of $S_{2n}$ is 0.975 MeV).

\begin{figure}[t]
	\centering
	\includegraphics[width=6.0cm]{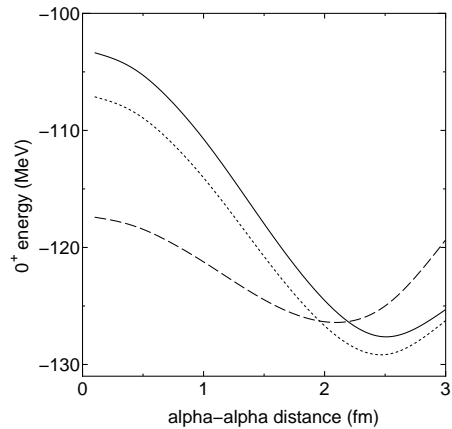} 
	\caption{.
The $0^+$ energy 
of $^{16}$O with a tetrahedron configuration of four $\alpha$'s
as a distance between $\alpha$ clusters.
The dotted line is the result of original F1 parameter set.
The results of F1'
parameter set is shown as solid line.
The dashed line is for the result calculated using
Volkov No.2 interaction~\cite{Volkov} with $M = 0.63$.
     }
\label{O16aa}
\end{figure}

For $^{16}$O, according to the $jj$-coupling shell model,
the ground state
corresponds to the closure of major shell ($p$ shell), where 
both spin-orbit attractive ($p_{3/2}$)  and repulsive ($p_{1/2}$) orbits
are filled and the contribution of the spin-orbit interaction cancels. 
Therefore, $\alpha$ breaking configurations are not expected to mix strongly, 
and here we introduce a four $\alpha$ model space,
which is known to coincide with the closed $p$ shell configuration
at the limit of relative distance between $\alpha$ clusters equal to zero. 
The $0^+$ energy of $^{16}$O
with a tetrahedron configuration of four $\alpha$'s
is shown in Fig.~\ref{O16aa}
as a function of the relative distance between $\alpha$ clusters.
The dotted line is the result of original F1 parameter set,
and the result of newly introduced F1' 
parameter set is shown as the solid line.
Here, F1' is designed to avoid small overbinding of $^{16}$O
when the original F1 parameter set is introduced, 
and the solid line is less attractive compared with
the dotted line by about $2 \sim 3$ MeV. 
The energies of $^{16}$O is obtained by superposing
the basis states with $\alpha$-$\alpha$ distances of 
0.1, 0.5, 1.0, 1.5, 2.0, 2.5, and 3.0 fm
based on GCM, and
the newly introduced parameter F1' parameter set gives $-126.9$ MeV
for the ground state
compared to the experimental value of $-127.619293$ MeV.

\begin{figure}[t]
	\centering
	\includegraphics[width=6.0cm]{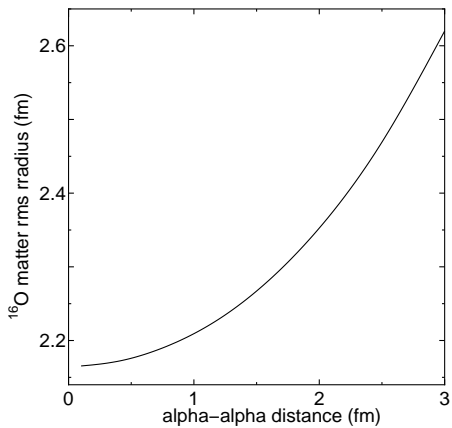} 
	\caption{.
The matter rms radius
for the $0^+$ state 
of $^{16}$O 
calculated with the tetrahedron configuration of four $\alpha$'s.
The horizontal axis shows
the relative distance between $\alpha$ clusters.
     }
\label{O16rms}
\end{figure}
\begin{figure}[t]
	\centering
	\includegraphics[width=6.0cm]{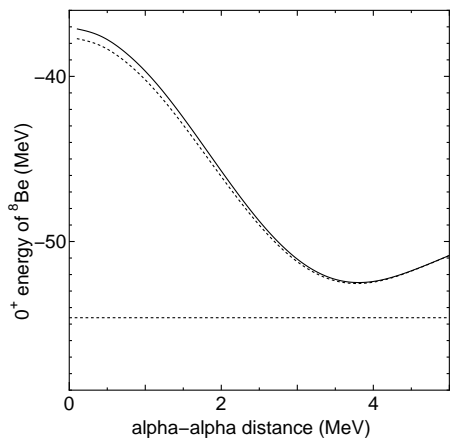} 
	\caption{
The $0^+$ energy curves of $^{8}$Be calculated with
$\alpha$+$\alpha$ model as a function of 
relative $\alpha$-$\alpha$ distance.
The dotted and solid lines are the results 
calculated with the original F1 parameter set
and newly introduced F1' parameter set.
The dotted line at $-54.61$ MeV shows
the threshold energy of $\alpha$+$\alpha$.
     }
\label{aa-en}
\end{figure}

The observed root mean square (rms) radius of $^{16}$O is quite large; the charge radius is
2.69 fm \cite{Angeli}, and we often underestimate the radius
by 0.1-0.2 fm, if we calculate with four $\alpha$ models
and use only two-body effective interaction 
such as Volkov interaction~\cite{Volkov}.
The dashed line in Fig.~\ref{O16aa} is for the result calculated using
Volkov No.2 interaction with $M = 0.63$.
In this case, the energy curve
is much shallower and feature of the curve is quite different from
the Tohsaki interaction cases.
The dashed line shows the energy minimum point 
around the $\alpha$-$\alpha$ distance of 2 fm.
Using the Tohsaki interaction with finite-range
three-body interaction terms, 
the solid line in Fig.~\ref{O16aa} shows that the lowest
energy is obtained with the $\alpha$-$\alpha$ distance of 2.5 fm,
larger than the result of Volkov interaction (dashed line) by 0.5 fm.
The matter rms radius
for the $0^+$ state 
of $^{16}$O with a tetrahedron configuration of four $\alpha$'s
 is shown in Fig.~\ref{O16rms} as a function of distances between $\alpha$ clusters.
When the $\alpha$-$\alpha$ distance is 2.5 fm,
which gives the lowest energy in the Tohsaki interaction cases,
the matter radius is 2.49 fm. 
This matter radius decreases to 2.35 fm
if the $\alpha$-$\alpha$ distance is 2.0 fm,
which gives the lowest energy in the Volkov interaction case.
The ground state of $^{16}$O obtained by superposing
the basis states with the $\alpha$-$\alpha$ distances of 
0.1, 0.5, 1.0, 1.5, 2.0, 2.5, and 3.0 fm
gives the rms matter radius of 2.49 fm (Tohsaki interaction F1' parameter set).
This value corresponds to the charge radius of 2.64 fm,
and the experimental value is almost reproduced.

Here we compare the F1 (original) and F1' (modified)
parameter sets of the
Tohsaki interaction. 
The $0^+$ energy curves of $\alpha$-$\alpha$ ($^8$Be)
calculated with F1 (dotted line) and F1' (solid line)
are compared in Fig.~\ref{aa-en}.
As explained previously, F1' is designed to avoid small overbinding of $^{16}$O
calculated with F1, and the solid line is slightly more repulsive at short
$\alpha$-$\alpha$ relative distances. However the difference is quite small 
and less than 1 MeV, and 
the character of the original F1 that the $\alpha$-$\alpha$ scattering phase shift is 
reproduced is not influenced by this modification.

For $^{12}$C, 
we prepare $jj$-coupling ($\alpha$ breaking) components of
the wave functions based on AQCM.
We introduce basis states with equilateral triangular configuration of 
three $\alpha$ clusters with the relative distances of
$R$ = 0.5, 1.0, 1.5, 2.0, 2.5, and 3.0 fm and change $\alpha$ clusters
to quasi clusters by giving dissolution parameter
$\Lambda$. The values of $\Lambda$ are chosen to be 0.2 and 0.4, since 
the states in between pure three $\alpha$ clusters ($\Lambda = 0$) and $jj$-coupling 
shell model limit ($\Lambda = 1$) 
are known to be important for the description of the ground state. 
In addition to these 12 basis states, 
we prepare 28 basis sates with various there $\alpha$ configurations 
by randomly generating Gaussian center parameters. 
This is because, the Hoyle state is a gas-like state without specific shape, 
and it has been known that not only equilateral triangular configuration,
various three $\alpha$ cluster configurations couple in this state. 

\begin{figure}[t]
	\centering
	\includegraphics[width=6.0cm]{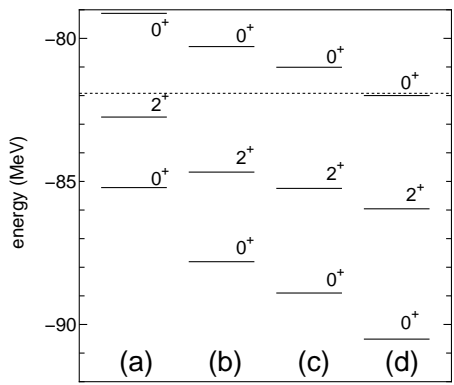} 
	\caption{
The energy levels of $^{12}$C.
Here (a) is the result without the spin-orbit interaction, and
(b), (c), and (d) show the results calculated with 1600 MeV, 1800 MeV, and 2000 MeV
for the strength of the spin-orbit terms of the G3RS interaction
($V_{ls}$ in Eq.~\ref{Vls}), respectively.
The dotted line at $-81.92$ MeV shows three $\alpha$ threshold energy.
     }
\label{c12-level}
\end{figure}

By superposing these 40 basis states based on GCM and diagonalizing the Hamiltonian,
energy eigen states are obtained.
The F1' parameter set of the Tohsaki interaction is adopted for the central part.
In Fig~\ref{c12-level}, the ground $0^+$, first $2^+$, and second $0^+$
states of $^{12}$C are shown 
together with the calculated three $\alpha$ threshold energy (dotted line).
The strength of the spin-orbit interaction,
$V_{ls}$ in Eq.~\ref{Vls}, is chosen as
$V_{ls} =$ 0 MeV, 1600 MeV, 1800 MeV, and 2000 MeV in (a), (b), (c), and (d), respectively.
The reasonable range of the strength of
$V_{ls} = 16000-2000$ MeV has been suggested
in the $^4$He+$n$
scattering phase shift analysis~\cite{Okabe}.
Without the spin-orbit interaction ($V_{ls} =$ 0 MeV),
the ground $0^+$ state of $^{12}$C is obtained at $-85.2$ MeV in Fig~\ref{c12-level} (a)
compared with the experimental value of $-92.161726$ MeV.
However, with the spin-orbit effect, the ground state
is obtained at $-87.8$ MeV in Fig~\ref{c12-level} (b) ($V_{ls} =$ 1600 MeV),
$-88.9$ MeV in (c) ($V_{ls} =$ 1800 MeV),
and $-90.5$ MeV in (d) ($V_{ls} =$ 2000 MeV).
Therefore, the absolute value of the binding energy of $^{12}$C 
can be almost reproduced with the present interaction and the model,
together with the
binding energies of $^4$He and $^{16}$O.
If we measure the energy from the three $\alpha$ threshold energy, the ground state 
of $^{12}$C is
$-3.3$ MeV in Fig~\ref{c12-level} (a), $-5.9$ MeV in (b), $-7.0$ MeV in (c), 
and $-8.6$ MeV in (d),
compared with the experimental value of $-7.2747$ MeV.
Thus the binding energy from the three $\alpha$ threshold is also reproduced
in the case of $V_{ls} = 1800$ MeV (Fig~\ref{c12-level} (c)).

The famous Hoyle state, the second $0^+$ state 
experimentally observed at $E_x = 7.65420$ MeV,
appears at 
$E_x = 6.1$ MeV in Fig~\ref{c12-level} (a),
$E_x = 7.5$ MeV in (b),
$E_x = 7.9$ MeV in (c),
and $E_x = 8.6$ MeV in (d).
If we measure from the three $\alpha$ threshold energy,
these energies correspond to
$E_x = 2.8$ MeV in Fig~\ref{c12-level} (a),
$E_x = 1.6$ MeV in (c),
$E_x = 0.9$ MeV in (c),
and $E_x = -0.1$ MeV in (d). 
Here again $V_{ls} = 1800$ MeV gives reasonable agreements with the experiment;
however the result implies that we need slightly larger number of basis states to reproduce Hoyle state
just above (experimentally 0.38 MeV) the three $\alpha$ threshold energy.

The traditional three $\alpha$ cluster models have a serious problem that
they give very small level spacing
between the ground $0^+$ state and the first $2^+$ state, which is the
first excited state of $^{12}$C~\cite{Fujiwara}.
In our case, $V_{ls} = 0$ MeV result  in Fig~\ref{c12-level} (a) shows that
the level spacing is 2.5 MeV and much
smaller compared with the experimental value of 4.4389131 MeV.
This is improved by the spin-orbit effect,
since the excitation from the ground $0^+$ state to the $2^+$ state
corresponds to one-particle one-hole excitation to a spin-orbit unfavored orbit
from the closure configuration
of spin-orbit attractive orbits in the $jj$-coupling shell model.
The $0^+-2^+$ level spacing becomes
3.1 MeV in (b), 3.7 MeV in (c), and 4.6 MeV in (c).
Similar trend has been reported in the recent antisymmetrized molecular dynamics~\cite{EnyoC}
and fermionic molecular dynamics~\cite{Neff} calculations.

The appearance of negative parity states in low-lying excitation energy has
been considered as the signature of the importance of
three $\alpha$ cluster structure in $^{12}$C.
Experimentally $3^-$ and $1^-$ states have been observed at 
$E_x = 9.6415$ MeV and 
$E_x = 10.84416$ MeV, respectively.
These $3^-$ and $1^-$ states are reproduced at
9.8 MeV and 12.6 MeV, respectively, 
when the strength of the spin-orbit interaction is chosen as
$V_{ls} = 1800$ MeV, which gives reasonable results for the $0^+$ and $2^+$ states.

\section{Summary}\label{summary}
We have tried to achieve consistent description of $^{12}$C and $^{16}$O,
which has been
a long standing problem of microscopic cluster model.
By taking into account the coupling with the $jj$-coupling shell model
and utilizing Tohsaki interaction, which is
finite-range three-body interaction, we have shown that
consistent understanding of these nuclei can be achieved.
The original Tohsaki interaction gives small overbound of about 3 MeV for $^{16}$O,
and this is improved by slightly modifying three-body Majorana exchange 
parameter. Also, so far the application of Tohsaki interaction has been limited
to $4N$ nuclei, and here, we added Bartlet and Heisenberg exchange terms
in the two-body interaction for the purpose of applying it to neutron-rich systems.

By applying Tohsaki interaction with finite-range
three-body interaction terms to $^{16}$O, the lowest
energy of the tetrahedron configuration of four $\alpha$'s
is obtained with very large $\alpha$-$\alpha$ distance (2.5 fm).
After performing GCM,
the ground state is obtained with the charge radius of 2.64 fm,
compared with the observed value of 2.69 fm. 
We often underestimate the radius
by 0.1-0.2 fm with four $\alpha$ models, if we calculate  
only within the two-body effective interactions, and this is significantly improved.

For $^{12}$C, we prepared various there $\alpha$ configurations 
by randomly generating Gaussian center parameters, and we mix
$jj$-coupling ($\alpha$ breaking) components based on AQCM.
The ground $0^+$ state of $^{12}$C is obtained at $-88.0 \sim -90.5$ MeV 
with reasonable strength of the spin-orbit interaction
compared with the experimental value of $-92.2$ MeV.
The absolute value of the binding energy of $^{12}$C (and also $^4$He and $^{16}$O)
can be almost reproduced with the present interaction and the model. 
If we measure the energy from the three $\alpha$ threshold energy,
the agreement with the experiment is even more reasonable.
The famous Hoyle state (second $0^+$ state) is reproduced just around
the three $\alpha$ threshold energy.
Also, traditional three $\alpha$ cluster models give very small level spacing
for the ground $0^+$ state and the first $2^+$ state, 
and this is significantly improved by the spin-orbit effect.
The appearance of negative parity states in low-lying excitation energy has
been considered as the signature of the three $\alpha$ cluster structure of $^{12}$C.
Experimentally $3^-$ and $1^-$ states have been observed at 
$E_x = 9.6415$ MeV and 
$E_x = 10.84416$ MeV, respectively, and these states are also
reproduced within the present framework.

\begin{acknowledgments}
The author thank the discussions with Prof. A. Tohsaki.
Numerical calculation has been performed at Yukawa Institute for Theoretical Physics,
Kyoto University. This work was supported by JSPS KAKENHI Grant Numbers 
716143900002.
\end{acknowledgments}


\begin{references}

\bibitem{Fujiwara}
Y. Fujiwara {\it et al.,}
Supple. of Prog. Theor. Phys. {\bf 68} 29 (1980).

\bibitem{Hoyle}
F. Hoyle, D. N. F Dunbar, W. A. Wenzel, W. Whaling, Phys. Rev. {\bf 92}, 1095c (1953).

\bibitem{Horiuchi}
Hisashi Horiuchi and Kiyomi Ikeda,
Prog. Theor. Phys. {\bf 40}, 277 (1968).

\bibitem{Descouvemont}
P. Descouvemont, Phys. Rev. C {\bf 47}, 210 (1993).

\bibitem{Tohsaki}
Akihiro Tohsaki, Phys. Rev. C {\bf 49}, 1814 (1994).

\bibitem{C-O}
Naoyuki Itagaki, Akira Ohnishi, and Kiyoshi Kat\=o,
Prog. Theor. Phys. {\bf 94}, 1019 (1995).


\bibitem{Simple}
	N. Itagaki, H. Masui, M. Ito, and S. Aoyama, Phys. Rev. C {\bf 71} 064307 (2005). 
\bibitem{Masui}
	H. Masui and N. Itagaki, Phys. Rev. C {\bf 75} 054309 (2007). 
\bibitem{Yoshida2}
	T. Yoshida, N. Itagaki, and T. Otsuka, Phys. Rev. C {\bf 79} 034308 (2009).
\bibitem{Ne-Mg}
	N. Itagaki, J. Cseh, and M. P{\l}oszajczak, Phys. Rev. C {\bf 83}, 014302 (2011).
\bibitem{Suhara}
  T. Suhara, N. Itagaki, J. Cseh, and M. P{\l}oszajczak,
Phys. Rev. C {\bf 87}, 054334 (2013).
\bibitem{Suhara2015}
Tadahiro Suhara and Yoshiko Kanada-En'yo,
Phys. Rev. C {\bf 91}, 024315 (2015).
\bibitem{Itagaki}
N. Itagaki, H. Matsuno, and T. Suhara, Prog. Theor. Exp. Phys., in press.
\bibitem{Volkov}
A.B. Volkov, Nucl. Phys. {\bf 74}, 33 (1965).

\bibitem{MV1}
T. Ando, K. Ikeda, and A. Tohsaki-Suzuki, Prog. Theor. Phys. {\bf 64}, 1608 (1980)

\bibitem{Gogny}
J. Decharge and D. Gogny, Phys. Rev. C {\bf 21}, 1568 (1980).

\bibitem{Enyo}
Y. Kanada-En`yo and Y. Akaishi, Phys. Rev. C {\bf 69}, 034306 (2004).



\bibitem{G3RS} 
R. Tamagaki, Prog. Theor. Phys. {\bf 39}, 91 (1968).



\bibitem{Okabe}
Shigeto Okabe and Yasuhisa Abe,
Prog. Theor. Phys. {\bf 61} 1049 (1979).


\bibitem{Brink}
D.M. Brink, in {\it Proceedings of the International School of Physics "Enrico Fermi" Course XXXVI},
edited by C. Bloch (Academic, New York, 1966), p. 247.

\bibitem{Angeli}
I. Angeli and K. P. Marinova, At. Data Nucl. Data Tables {\bf 99}, 69 (2013).



\bibitem{EnyoC}
	Y. Kanada-En'yo, Phys. Rev. Lett. \textbf{81}, 5291 (1998);
	Y. Kanada-En'yo, Prog. Theor. Phys. \textbf{117}, 655 (2007).
\bibitem{Neff}
	T. Neff and H. Feldmeier, Nucl. Phys. \textbf{A738}, 357 (2004).
	


\end{references}
\end{document}